\documentclass[letterpaper,10pt,onecolumn,oneside,final,journal]{IEEEtran}      % Use this line for a4
\IEEEoverridecommandlockouts                              % This command is only
   \usepackage{graphicx}
\usepackage[tight,footnotesize]{subfigure}
\usepackage{stfloats}
\usepackage{mathptmx} % assumes new font selection scheme installed
\usepackage{amsmath} % assumes amsmath package installed
\usepackage{amssymb}  % assumes amsmath package installed
\newcounter{mytempeqncnt}
\title{Disclaimer: This work has been accepted for publication in the IEEE Antennas and Wireless Propagation Letter. Copyright with IEEE. Personal use of this material is permitted. However, permission to reprint/republish this material for advertising or promotional purposes or for creating new collective works for resale or redistribution to servers or lists, or to reuse any copyrighted component of this work in other works must be obtained from the IEEE. This material is presented to ensure timely dissemination of scholarly and technical work. Copyright and all rights therein are retained by authors or by other copyright holders. All persons copying this information are expected to adhere to the terms and constraints invoked by each author's copyright. In most cases, these works may not be reposted without the explicit permission of the copyright holder.
For more details, see the IEEE Copyright Policy.\\
\vspace{10em}
An Analytical Link Loss Model for On-Body Propagation Around the Body Based on Elliptical Approximation of the Torso with Arms' Influence Included
}

\author{R. Chandra and A. J Johansson 
\thanks{R. Chandra and A. J Johansson are at Department of Electrical and Information Technology, Lund University, Lund, Sweden. This work was supported by the grant from SSF, Sweden for Ultra-Portable Devices (UPD) for wireless communication project.}%
\thanks{E-mail: rohit.chandra@eit.lth.se}
}
\begin{document}

\maketitle
\thispagestyle{empty}
\pagestyle{empty}

%%%%%%%%%%%%%%%%%%%%%%%%%%%%%%%%%%%%%%%%%%%%%%%%%%%%%%%%%%%%%%%%%%%%%%%%%%%%%%%%
\begin{abstract}
An analytical model for estimating the link loss for the on-body wave propagation around the torso is presented. The model is based on the attenuation of the creeping waves over an elliptical approximation of the human torso and includes the influence of the arms. The importance of including the arms' effect for a proper estimation of the link loss is discussed.
The model is validated by the full-wave electromagnetic simulations on a numerical phantom. %A good agreement is obtained between the model and the simulations.  %The usability of the model lies in the fact of its estimating the link loss between wireless body area network devices around the torso in a very short time as compared to the memory and the time consuming full-wave computer simulations. 
\end{abstract}


\section{Introduction}
Wireless Body Area Network (WBAN) has emerged as a key technology in health care and consumer electronics~\cite{latre}. The devices in WBAN can be broadly divided into two categories: (a) wearable or on-body (b) implantable. The communication between two wearable/on-body devices located on the opposite side of the body is through creeping waves~\cite{chandra1}-\cite{chandra2}. Creeping waves undergoes exponential attenuation with the distance~\cite{ryk}. Hence, the estimation of the link loss between the on-body devices is essential for the link budget and deciding the sensitivity for a reliable wireless link. Statistical and deterministic propagation/link loss models for various WBAN scenarios are presented in~\cite{ryk}-\cite{cotton}. The statistical approach models the link loss in a dynamic scenario when the body is moving~\cite{cotton} and the deterministic approach models the link loss in a stationary environment when the body is static~\cite{alves}. The deterministic link loss is discussed in this letter.  

The importance of the torso shape for correctly quantifying the path-loss around the body is discussed in~\cite{eid}. An analytical model for the creeping wave propagation around the body based on a circular approximation of the cross-section of the torso is presented in~\cite{alves} and for an elliptical approximation of the torso is presented in~\cite{chandra2}. In~\cite{chandra2}, it is shown that a circular cross-section under-estimates the link loss. The effects due to the arms are excluded in the models presented in~\cite{alves} and~\cite{chandra2}. However, in~\cite{chandra3}, we have shown that there is a significant influence on the link loss due to the reflections from the arms. Usually, the arms are present at the side of the torso and will influence the deterministic link loss. Hence, a model which is easier to handle than time consuming simulations is needed to evaluate the effects of the arms on the wave propagation around the torso for designing a reliable link. The goal of this letter is to develop such a model by extending the analytical model presented in~\cite{chandra2} to include the effects of the arms. The developed model is validated through the full-wave FDTD simulations done in SEMCAD-X~\cite{semcad} in $2.45$ GHz ISM band. %Six different scenarios with different arm positions are considered. Good agreement with the model is obtained.

   \begin{figure}[thpb]
      \centering
      \includegraphics[scale=.135]{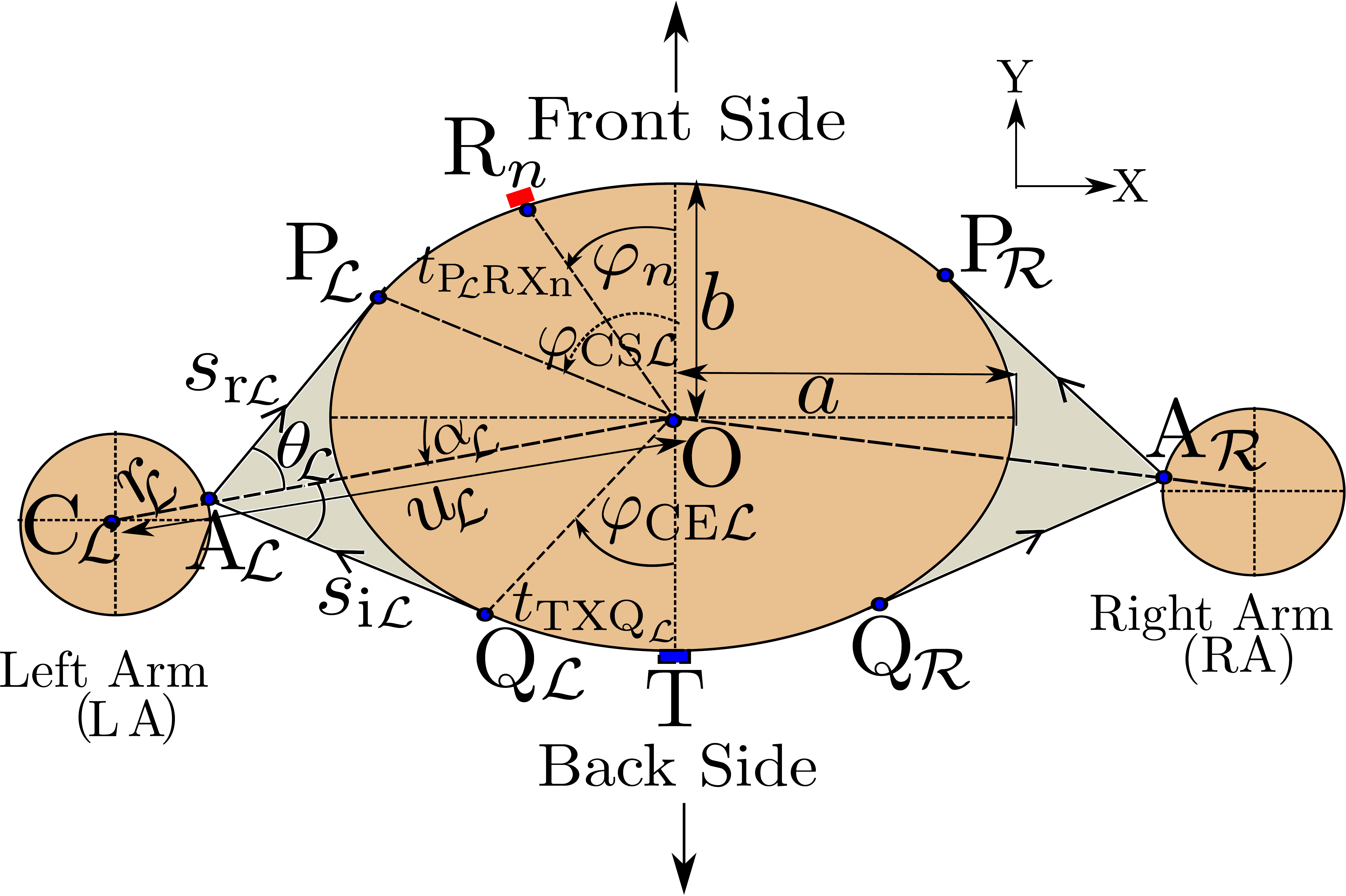}
      \caption{Cross-sectional view of the torso at the level of the antennas showing the parameters and coordinates. The parameters and the coordinates are described in Table~\ref{paramter_table} }%
     \label{arm_reflection}
     \vspace{-10pt}
   \end{figure}
%-------------------- Table Begin------------------------------
\begin{table*}[ht]
\caption{Description of the Parameters for the Analytical Model} % title of Table
\centering % used for centering table
\scriptsize
\begin{tabular}{l l} % centered columns (4 columns)
\hline\hline %inserts double horizontal lines
Parameter/Coordinates & Description \\ [0.5ex] % inserts table
%heading
\hline 
$a$; $b$ & semi-major axis of the ellipse; semi-minor axis of the ellipse  \\ % inserting body of the table
%$b$ & semi-minor axis of the ellipse \\
$p$ & perimeter of the ellipse (not shown in Fig.~\ref{arm_reflection}) \\
O(0,0) & Origin and center of the ellipse\\
$\mathrm{R}_{n}(x_{\mathrm{RXn}}, y_{\mathrm{RXn}})$ & coordinates of the Rx at $\mathrm{n^{th}}$ position\\
$\mathrm{T}(x_{\mathrm{TX}},y_{\mathrm{TX}})$ & coordinates of the Tx\\
$d_{\mathrm{n}}$ & anti-clockwise distance over ellipse from the Tx to $\mathrm{R}_n$ (not shown in Fig.~\ref{arm_reflection})\\
$\varphi_{\mathrm{n}}$ & angle of the Rx at $\mathrm{n^{th}}$ position from the semi-minor axis on front side of body\\
$\mathrm{C}(x_{\mathrm{AC}}, y_{\mathrm{AC}})$ & coordinates of the center of arm\\
$u$ & distance between center of ellipse and arm\\
$r$ & radius of the arm\\
$\alpha$ & angle between semi-major axis of ellipse and line joining center of arm with ellipse'e center\\
$\mathrm{Q}(x_{\mathrm{CE}}, y_{\mathrm{CE}})$ & point of leave on ellipse where creeping ends\\ 
$t_{\mathrm{TXQ}}$ & distance over ellipse from Tx to the point of leave\\
$\varphi_{\mathrm{CE}}$ & angle between the TX and the point where creeping ends\\
$\mathrm{A}(x_{\mathrm{A}}, y_{\mathrm{A}})$ & point of reflection on arm\\
$s_{\mathrm{i}}$ & distance between point of leave on ellipse and point of reflection on arm\\
$\mathrm{P}(x_{\mathrm{CS}}, y_{\mathrm{CS}})$ & point of contact on ellipse where creeping again starts\\
$s_{\mathrm{r}}$ & distance between point of reflection on arm and point of contact \\
$\theta$ & angle of incidence/angle of reflection\\
$t_{\mathrm{PRXn}}$ & distance over ellipse from point of contact to Rx at $\mathrm{n^{th}}$ position\\ 
$\varphi_\mathrm{CS}$ & angle of the point of contact from semi-minor axis on front side of body \\
\hline %inserts single line
%\smallsize{*not shown in Fig.~\ref{arm_reflection}}& \\
%\smallsize{'L' and 'R' in subscript in~Fig.\ref{arm_reflection} represents left side and right side resp. for the parameters shown here &\\
\vspace{-15pt}
\end{tabular}
\label{paramter_table} % is used to refer this table in the text
\end{table*}   
%-------------------- Table End------------------------------ 
%\vspace{-12pt}
%%%%%%%%%%%%%%%%%%%%%%%%%%%%%%%%%%%%%%%%%%%%%%%%%%%%%%%%%%%%%%%%%%%%%%%%%%%%%%%%
\section{Derivation of the Analytical Model}

%The presence of the arms at the side of the torso results in the reflection of the electromagnetic (EM) waves which leaves the torso after creeping. The reflected waves then add up at the receiver with the clockwise and the anti-clockwise creeping waves, constructively or destructively resulting in different received power level than the one without the arms. The effect of the interference is more prominent when the receiver and the transmitter are on the opposite side. Moreover, the influence of the arms is significant when they are close to the torso.

%In this section, the analytical channel model for the link loss around the torso presented in~\cite{chandra2} is extended to include the influence of the arms. 
Let us consider a case shown in Fig.~\ref{arm_reflection}. The transmitter is fixed at the back side of the torso and the receiver is moved along the front side. The torso is shown by an ellipse and the arms are approximated by circle. In Fig.~\ref{arm_reflection}, $\mathcal{L}$ in the subscript is used for the left side of the torso and $\mathcal{R}$ for the right side of the torso. The coordinates and the parameters shown in Fig.~\ref{arm_reflection} are described in Table~\ref{paramter_table}. In the presence of the arms at the side of the torso, the electromagnetic (EM) waves from the transmit antenna (Tx) reaches the receive antenna (Rx) by two ways: (a) creeping wave starting at the Tx and then reaching the Rx through a clockwise and an anti-clockwise path around the torso and (b) waves leaving from the Tx directly or creeping for some distance over the torso and then leaving the torso tangentially at the point of leave (Q), getting reflected by the arms and reaching the Rx directly or after creeping from the point of contact (P).
\vspace{-5pt}
   
\subsection{Calculation of Parameters and Coordinates}
The parameters which should be known are $a$, $b$, $r$, coordinates of the center of the arms and the position of the transmitter. All other parameters and coordinates are calculated from the coordinate geometry using these parameters.

The angle $\alpha$ is calculated as: $\alpha = \mathrm{tan}^{-1}(|y_{\mathrm{AC}}|/|x_{\mathrm{AC}}|)$. %\begin{equation}
The x-coordinate of the point of reflection $\mathrm{A}$ is $x_{\mathrm{A}} = (u-r)\mathrm{cos\alpha}$ and the y-coordinate is $y_{\mathrm{A}}=(u-r)\mathrm{sin\alpha}$, where $u = \sqrt{x_{\mathrm{AC}}^2 + y_{\mathrm{AC}}^2}$. It should be noted that proper sign of the coordinates should be considered depending on the quadrant in which the arms are present. The tangents to the ellipse from the point of reflection $\mathrm{A}(x_{\mathrm{A}}, y_{\mathrm{A}})$ will have slopes, $m \in \left\{m_1,m_2 \right\}$ given by:
\begin{equation}
\label{slope}
m = \frac{x_{\mathrm{A}}y_{\mathrm{A}} \pm \sqrt{x_{\mathrm{A}}^2b^2 + y_{\mathrm{A}}^2a^2 - a^2b^2}}{x_{\mathrm{A}}^2 - a^2}
\end{equation}
The coordinates of the point of leave, $\mathrm{Q}(x_{\mathrm{CE}}, y_{\mathrm{CE}})$ and the point of contact, $\mathrm{P}(x_{\mathrm{CS}}, y_{\mathrm{CS}})$ is calculated by solving the tangents with the ellipse. $s_i$ and $s_r$ can then be calculated by distance formula between two points. With the knowledge of $m_1$ and $m_2$, angle between the tangents can be calculated. The angle of incidence/reflection, $\theta$, is half of the angle between the tangents. 

Any angle $\varphi$ between a line joining a point with x-coordinate $x$ on an ellipse to the center and the minor axis can be calculated by:
\begin{equation}
\label{angle}
\varphi = \mathrm{tan}^{-1}\left(\frac{a}{b}\frac{x}{\sqrt{a^2 - x^2}}\right)
\end{equation}
From~(\ref{angle}), angles $\varphi_{\mathrm{n}}$, $\varphi_{\mathrm{CS}}$ and $\varphi_{\mathrm{CE}}$ can be calculated. The arc length $t$ of an ellipse between any two angles $\varphi_i$ and $\varphi_j$ is given by~\cite{balanis}:
\begin{equation}
\label{arc_length}
t = ab\int_{\varphi_i}^{\varphi_j}\frac{(a^4\cos^2\varphi+ b^4\sin^2\varphi)^\frac{1}{2}}{(a^2\cos^2\varphi+ b^2\sin^2\varphi)^\frac{3}{2}}\mathrm{d}\varphi
\end{equation}
Using~(\ref{arc_length}) and integrating between proper values of $\varphi_i$ and $\varphi_j$, $p$, $d_{\mathrm{n}}$, $t_\mathrm{TXQ}$ and $t_\mathrm{PRXn}$ can be calculated. For example, $p$ can be calculated by integrating the expression in~(\ref{arc_length}) between $0$ and $2{\pi}$ and $t_\mathrm{PRXn}$ between $\varphi_{\mathrm{n}}$ and $\varphi_{\mathrm{CS}}$.
\vspace{-7pt}

\begin{figure*}[!b]
% ensure that we have normalsize text
%\vspace*{1pt}
\hrulefill
\normalsize
% Store the current equation number.
\setcounter{mytempeqncnt}{\value{equation}}
% Set the equation number to one less than the one
% desired for the first equation here.
% The value here will have to changed if equations
% are added or removed prior to the place these
% equations are referenced in the main text.
\setcounter{equation}{7}
\begin{eqnarray}
\label{rx_pwr_elliptical_reflection}
{LL_n|}_{\mathrm{dB}} &=& -10\mathrm{log_{10}}\left[\frac{G_{RX}G_{TX}\lambda^2}{4\pi^2}\left(\vline \frac{e^{-{L_{ac}}_n}}{d_n}e^{-jkd_n}+ \frac{e^{-{L_{c}}_n}}{p-d_n}e^{-jk(p-d_n)} \right.\right. \nonumber \\ 
&&\left. \left. + \sum_{\mathcal{J}=\mathcal{L}}^{\mathcal{R}}\frac{\rho}{2\sqrt{2}}\frac{e^{-L_\mathrm{TQ_\mathcal{J}}} e^{-L_\mathrm{P_\mathcal{J}RX_n}}\mathrm{cos}\gamma_\mathcal{J}}{t_\mathrm{TXQ_\mathcal{J}}+s_\mathrm{i\mathcal{J}}+s_\mathrm{r\mathcal{J}}+t_\mathrm{P_\mathcal{J}RXn}}e^{-jk(t_\mathrm{TXQ_\mathcal{J}}+s_\mathrm{i\mathcal{J}}+s_\mathrm{r\mathcal{J}}+t_\mathrm{P_\mathcal{J}RXn})} \vline \right)^2 \right]
\end{eqnarray}
\setcounter{equation}{\value{mytempeqncnt}}
\end{figure*}

\subsection{Link Loss Model} 
The model is applicable for the attenuation of the vertical component of the creeping wave's electric field over a conducting elliptical path as discussed in~\cite{chandra2}. The complex attenuation $L(\varphi_i,\varphi_j)$ over a conducting elliptical path between angles, $\varphi_i$ and $\varphi_j$, for the vertical component of the electric field is~\cite{balanis}:
\begin{eqnarray}\label{l}
L(\varphi_1,\varphi_2) = \frac{(k)^{1/3}}{2}\left(\frac{3\pi ab}{4} \right)^{2/3}e^{\frac{j\pi}{6}} &&\nonumber \\
.\int_{\varphi_1}^{\varphi_2}\frac{ab}{\sqrt{[a^4\cos^2\varphi + b^4\sin^2\varphi][a^2\cos^2\varphi + b^2\sin^2\varphi]}}\mathrm{d}\varphi
\end{eqnarray}
where $k$ is the wave number in a free space. With arms at the side of the torso, there will be two additional paths apart from the clockwise and the anti-clockwise creeping path that will contribute to the received power as shown in Fig.~\ref{arm_reflection}. The first path $TQ_\mathcal{L}\rightarrow Q_\mathcal{L}A_\mathcal{L}\rightarrow A_\mathcal{L}P_\mathcal{L}\rightarrow P_\mathcal{L}R_n$, is the path of the EM wave which creeps over the left side of the torso for a distance $t_\mathrm{TXQ_\mathcal{L}}$, then leaves the torso tangentially at the point of leave $Q_\mathcal{L}$, travel in a free space for a distance $s_\mathrm{i}$, gets reflected by the left arm at the point of reflection $A_\mathcal{L}$, travel in a free space for a distance $s_\mathrm{j}$ and then creeps to the receiver at $R_n$ for a distance $t_\mathrm{P_\mathcal{L}RXn}$ from the point of contact $P_\mathcal{L}$. The second path is a similar path, $TQ_\mathcal{R}\rightarrow Q_\mathcal{R}A_\mathcal{R}\rightarrow A_\mathcal{R}P_\mathcal{R}\rightarrow P_\mathcal{R}R_n$ on the right side. To keep the model simple, it is assumed that the reflected wave after contacting the torso do not interfere with the clockwise and the anti-clockwise creeping waves, rather it creeps towards the receive antenna from the point of contact. The total electric field at the receiver is given by the sum of the electric field of the waves from the four paths:
${\bf E} = {\bf E}_{c} + {\bf E}_{ac} + {\bf E}_{TA_\mathcal{L}P_\mathcal{L}R_n} + {\bf E}_{TA_\mathcal{R}P_\mathcal{R}R_n}$. For the clockwise (subscript c) and the anti-clockwise (subscript ac) creeping waves, ${\bf E}_{i} = {\bf E}_{0i}e^{-L_i}$ where the subscript $i$ can be $c$ or $ac$. $L_i$ can be calculated from (\ref{l}) with proper values of $\varphi$ for different receiver positions.  
The reference electric field ${\bf E}_{0i}$ at a distance $z_i$ from the transmitting antenna on a conducting surface is given by \cite{alves}:
\begin{equation}
{\bf E}_{0i} = 2\sqrt{\frac{\eta_0}{2\pi}}\frac{\sqrt{P_{TX}G_{TX}}}{z_i}\mathrm{e}^{-jkz_i}
\end{equation}
where $\eta_0$ is the wave-impedance in a free space and $z_i = p-d_\mathrm{n}$ for $i = c$ and $z_i = d_\mathrm{n}$ for $i = ac$.
The received electric field for the reflected wave from the arm is modeled as ${\bf E}_{TA_\mathcal{J}P_\mathcal{J}R_n} = {\bf E}_{0\mathcal{J}}e^{-L_\mathrm{TQ_\mathcal{J}}}e^{-L_\mathrm{P_\mathcal{J}RXn}}$ for $\mathcal{J} = \mathcal{L}$ or $\mathcal{R}$. $L_\mathrm{TQ_\mathcal{J}}$ is the attenuation factor from the Tx to the point of leave $Q_\mathcal{J}$ and $L_\mathrm{P_{\mathcal{J}}RXn}$ is attenuation factor from the point of contact $\mathrm{P_\mathcal{J}}$ to the Rx. The reference electric field for the reflected wave is given by~\cite{david}:
\begin{equation}
{\bf E}_{0\mathcal{J}} = \rho\frac{1}{\sqrt{2}}\sqrt{\frac{\eta_0}{2\pi}}\frac{\sqrt{P_{TX}G_{TX}}}{z_\mathcal{J}}\mathrm{e}^{-jkz_\mathcal{J}}
\end{equation}
where $z_\mathcal{J} = t_\mathrm{TXQ_\mathcal{J}} + s_\mathrm{i\mathcal{J}} + s_\mathrm{r\mathcal{J}}+  t_\mathrm{P_\mathcal{J}RXn} $ for $j = \mathcal{L}$ or $\mathcal{R}$ and $\rho$ is the reflection coefficient of the arm at an angle of incidence $\theta$~\cite{molisch}.

The received power is $P_{RX} = \frac{|{\bf E}|^2 A_{RX}}{2\eta_0}$ where $A_{RX} = G_{RX}\frac{\lambda^2}{4\pi}$ is the receive antenna aperture ($G_{RX}$ is the gain of the antenna and $\lambda$ is the wavelength in a free space). Substituting ${\bf E} = {\bf E}_{c} + {\bf E}_{ac} + {\bf E}_{TA_\mathcal{L}P_\mathcal{L}R_n} + {\bf E}_{TA_\mathcal{R}P_\mathcal{R}R_n}$ in the received power's expression, the link loss, ${LL_n|}_{\mathrm{dB}}$ at the $n^{th}$ receiver position can be written as:
\begin{equation}
\label{rx_pwr_elliptical_reflection_db}
{LL_n|}_{\mathrm{dB}} =-10\mathrm{log_{10}}{\frac{P_{RX}}{P_{TX}}\vline}_n 
\end{equation} 
The detail equation of the link loss is shown in (\ref{rx_pwr_elliptical_reflection}). The positions of the receive antenna which lies between the point of leave, $\mathrm{Q}_\mathcal{J}$ and the point of contact $\mathrm{P}_\mathcal{J}$ for $\mathcal{J}$ $=$ $\mathcal{L}$ or $\mathcal{R}$, receives the reflected wave from the arm directly. These positions lies on the ellipse within the shaded portion between the arm and the ellipse shown in~Fig.~\ref{arm_reflection}. However, if the direct received wave is in the direction of the null of the receive antenna, it will not contribute to the received power. This may be a case as the antennas for the on-body propagation are usually designed to minimize the power in the direction away from the body. To take care of such a case, a factor cos$\gamma$ is multiplied to the received reflected field, where $\gamma = 0$ if $R_n$ lies between $\mathrm{P}_\mathcal{L}$ and $\mathrm{P}_\mathcal{R}$, else it is equal to the angle between the tangent to the ellipse at the receiver position and the reflected wave. Similarly, if the transmit antenna is placed at a position where $|x_\mathrm{TX}| > |x_\mathrm{CE\mathcal{J}}|$ for $\mathcal{J}$ $=$ $\mathcal{L}$ or $\mathcal{R}$, the incident wave will be directly received by the arm. 

\begin{figure*}[thpb]
      \centering
      
    			\subfigure[]{\label{casea}\includegraphics[scale=0.17]{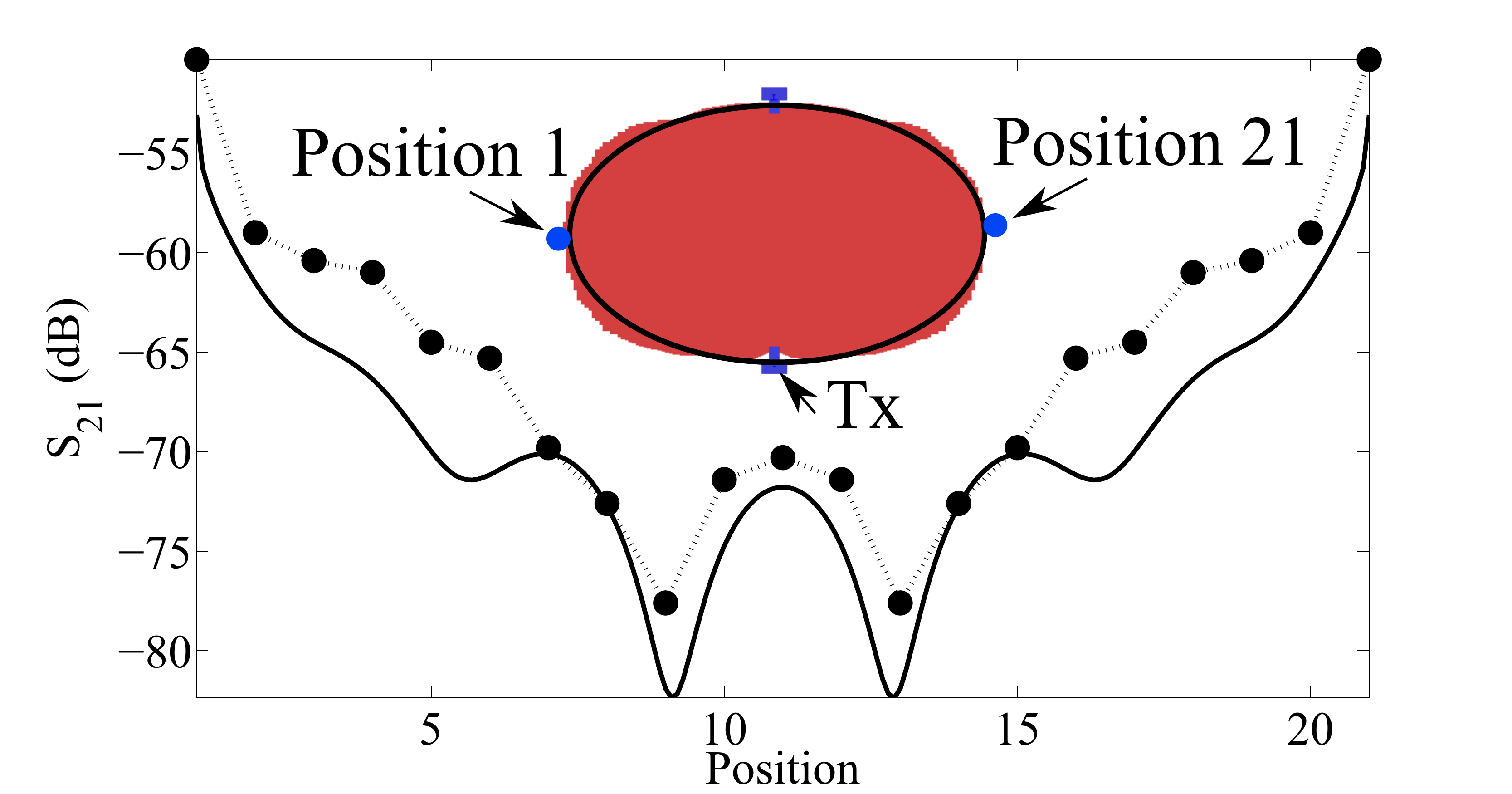}} 
    			\subfigure[]{\label{caseb}\includegraphics[scale=0.17]{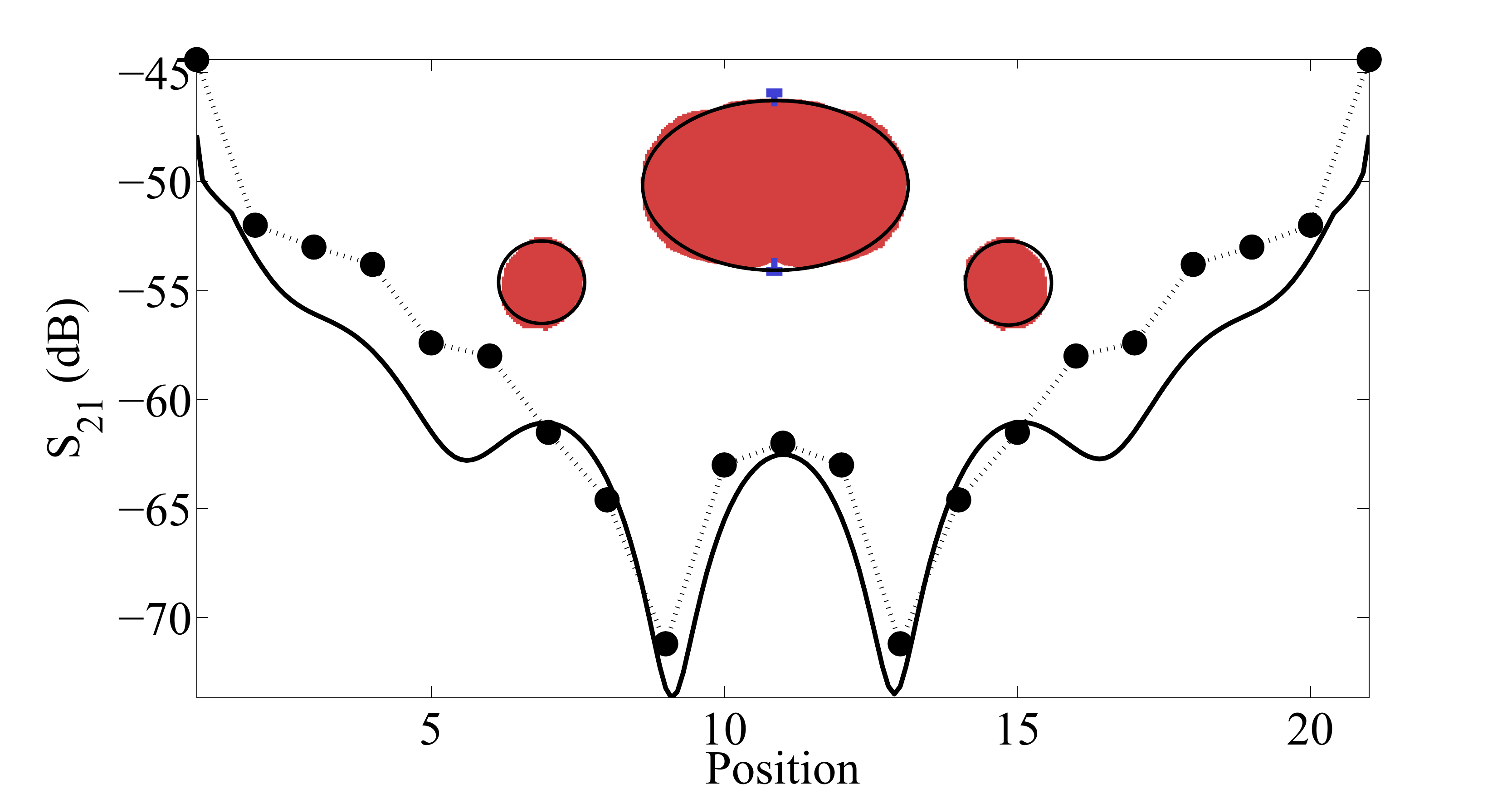}}
    			\subfigure[]{\label{casec}\includegraphics[scale=0.17]{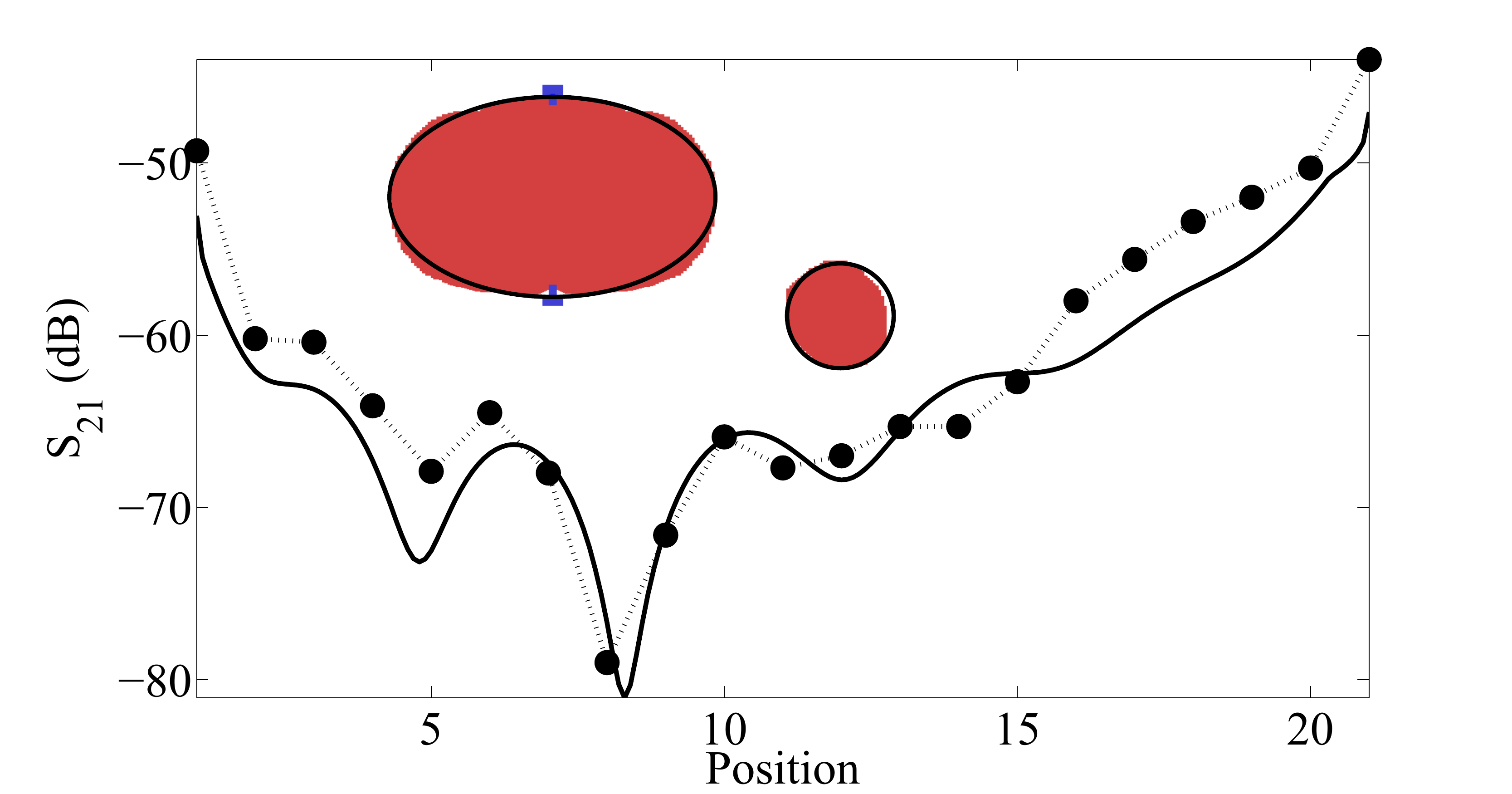}}\\
    			\subfigure[]{\label{cased}\includegraphics[scale=0.17]{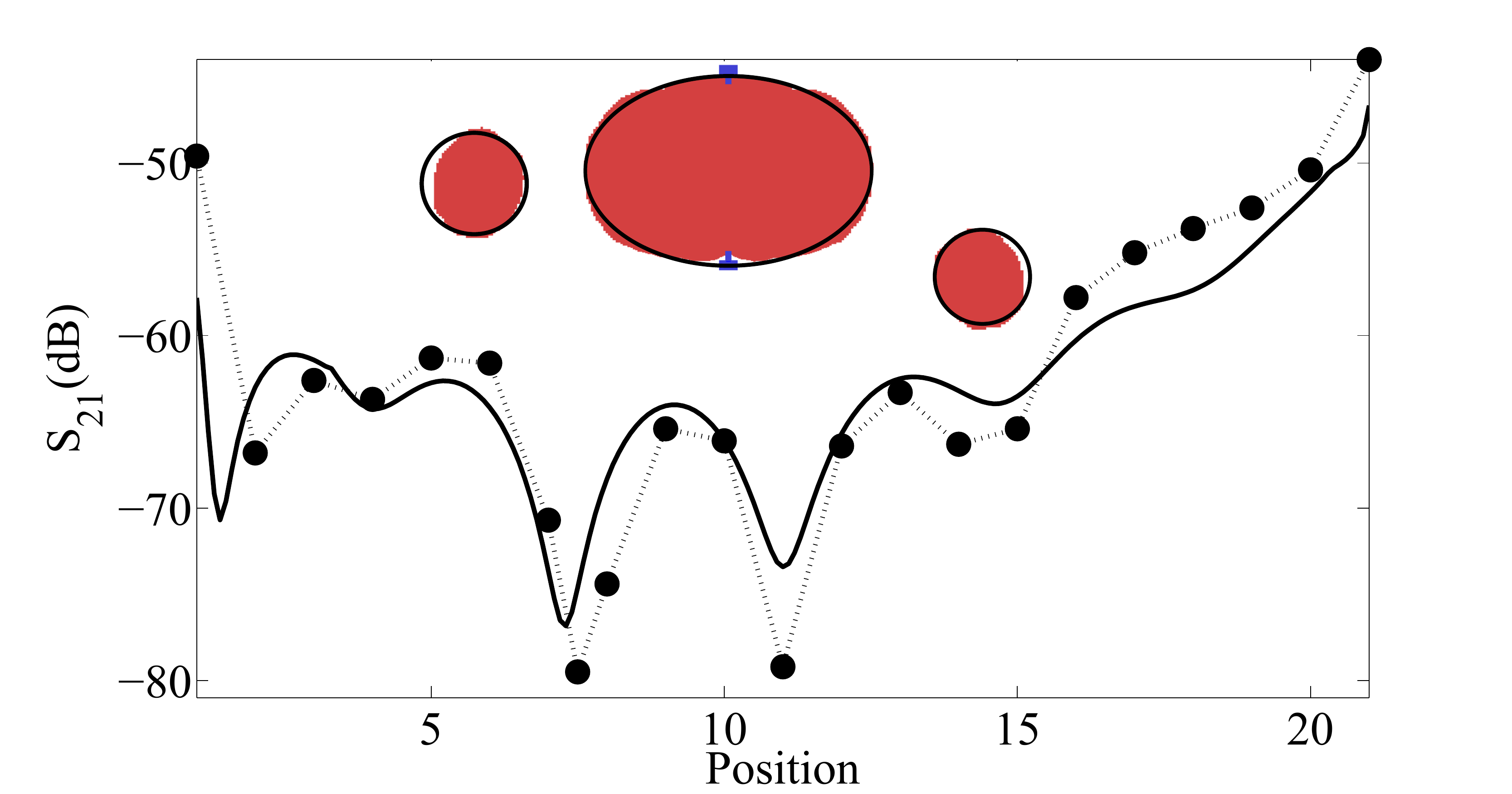}}
    			\subfigure[]{\label{casee}\includegraphics[scale=0.17]{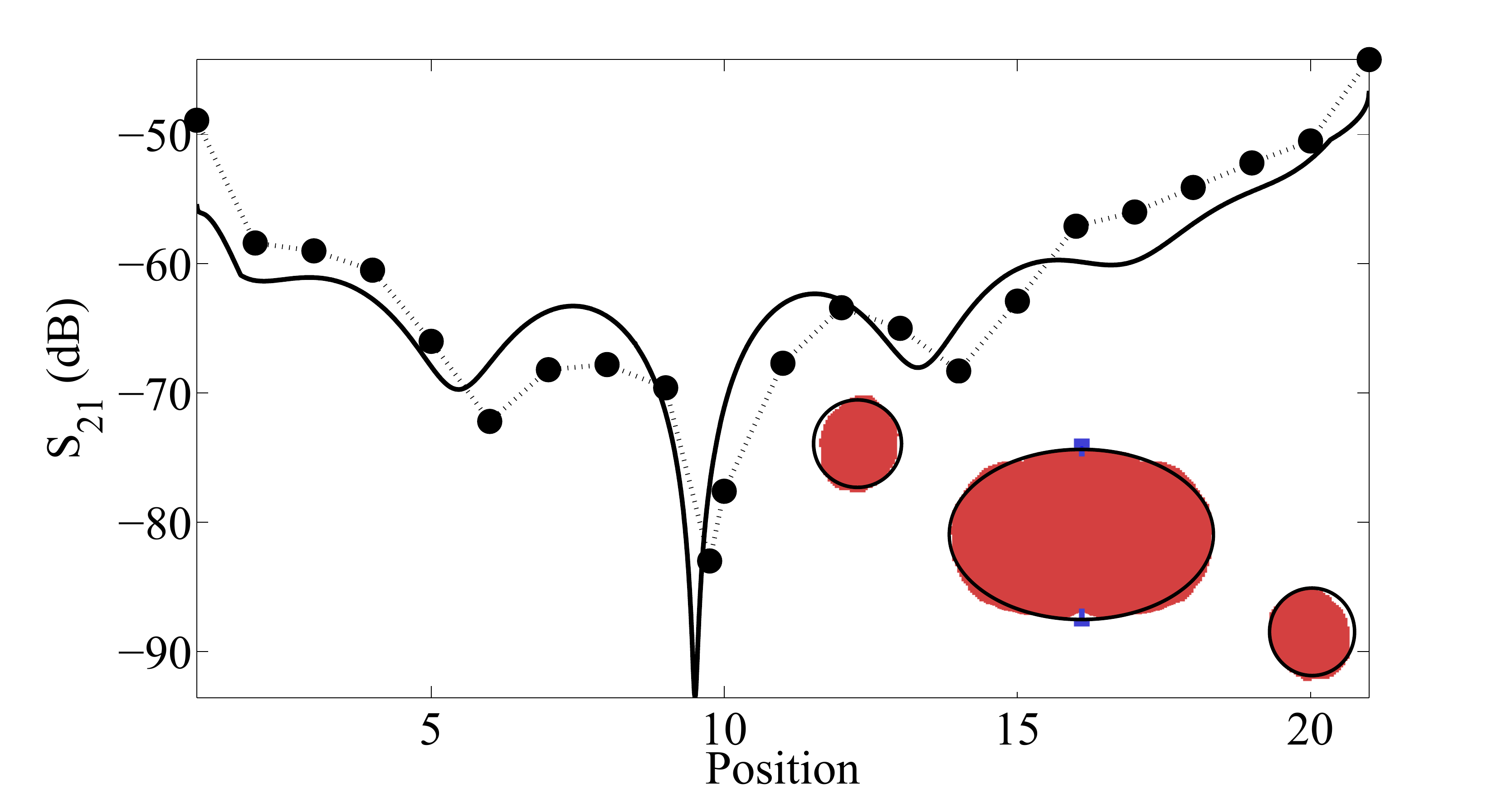}}
    			\subfigure[]{\label{casef}\includegraphics[scale=0.17]{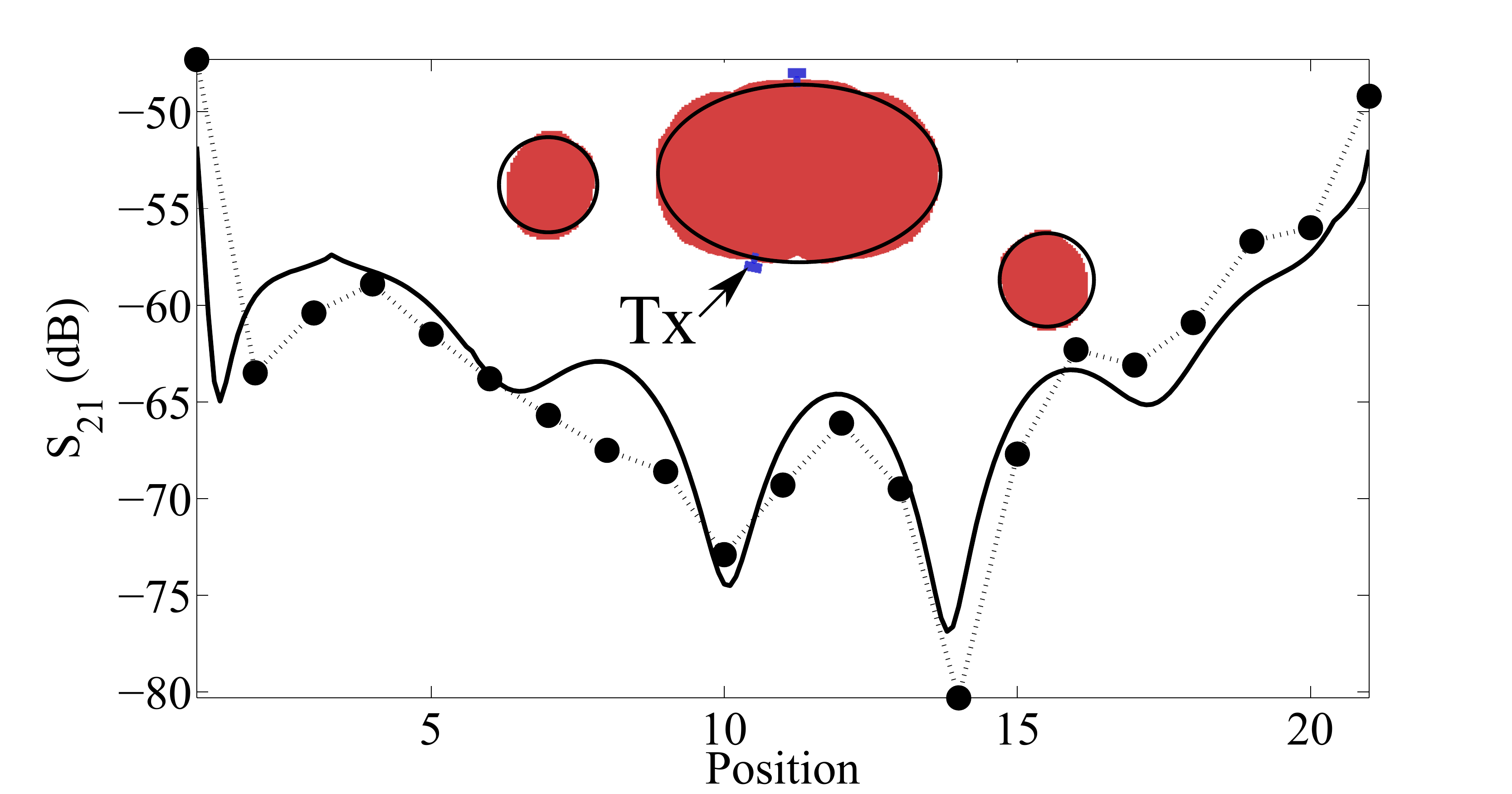}}
    			\caption{Simulated (dotted line) $S_{21}$ versus Analytical (solid line) $S_{21}$. Arms are not included in (a). In (a)-(e), the transmit antenna is at the central back position whereas in (f) the transmit antenna is shifted from the central back position towards left by $43$ mm.}
    			\label{results}  
\vspace{-5pt}    			    
\end{figure*} 

\section{Validation of the Analytical Model} 
The validation of the model is done over a truncated numerical phantom~\cite{chandra3} with homogeneous electrical properties of muscle (permittivity = $52$, conductivity = $1.7$ S/m). The phantom has dimensions of a typical adult male with $a = 144$ mm, $b = 93.6$ mm and $r = 50$ mm. These parameters in practice can be obtained through measurement of the user's torso. The measured waist-to-waist value will be equal to $2a$ and the measured abdomen-to-back will be equal to $2b$. $r$ could be calculated by measuring the perimeter of the arm and equating it with the perimeter of a circle with radius $r$. A truncated phantom is used as the whole body has a minimal influence on the link around the torso~\cite{chandra3}. The validation is done through the full-wave electromagnetic simulations in SEMCAD-X which uses the FDTD method. If the validation is done through measurement, post processing on the measured data is required because of the effects like leakage current from the cable, ground reflections and change in the path-length due to respiration and body movements. Moreover, due to the finite size of the antenna the detection of fading dips is not possible~\cite{alves}. Hence, full-wave simulation is chosen. 

Six different scenarios are considered for the validation of the analytical model. In all these six scenarios, the receiver antenna is moved along the front side of the abdomen at $21$ positions at a spacing of $x = a/10$. Additionally, few more positions are considered between these $21$ positions to confirm the fading dips. These scenarios are shown in the inset of Fig.~\ref{results} which also shows the plots for the simulated and the analytical $S_{21}$. The link loss ($LL$) in (\ref{rx_pwr_elliptical_reflection}) is the negative of $S_{21}$ in dB. It could be seen that a good agreement between the simulations and the analytical model is obtained. Some differences might occur due to the fact that the torso is not completely elliptical in shape. Variations in the antenna gain at different positions may also contribute to these differences~\cite{chandra2}. The antenna used as the receiver and the transmitter is a small monopole antenna matched in $2.45$ GHz ISM band~\cite{chandra1} which is vertically polarized (w.r.t. the body). The value of the gain used in the analytical equation is $-7.3$ dBi which is the gain of the antenna in the direction of the creeping wave at the central abdomen position. More discussion about the usage of the gain of the on-body antenna for the creeping wave can be found in~\cite{alves},~\cite{chandra2}.

\section{Evaluation of the Arms' Effect}
It is critical to include the effect of the arms while estimating the link loss because the reflected waves from the arms at a intended receiver position might interfere destructively with the on-body creeping waves. Hence, the link loss at that receiver position will be higher than the link loss obtained without considering the effect of the arms. For eg., let us consider a case when the receiver is placed midway between position $9$ and position $10$. The link loss is about $78$ dB without considering the effects of the arms (Fig.~\ref{casea}) whereas it is about $94$ dB for the same receiver position with the arms as in Fig.~\ref{casee}. Hence, the signal reception at this position will be poor if the receiver is designed to handle $80$ dB of the link loss considering the loss without the arms. 

\begin{figure}[thpb]
      \centering
      \includegraphics[scale=.175]{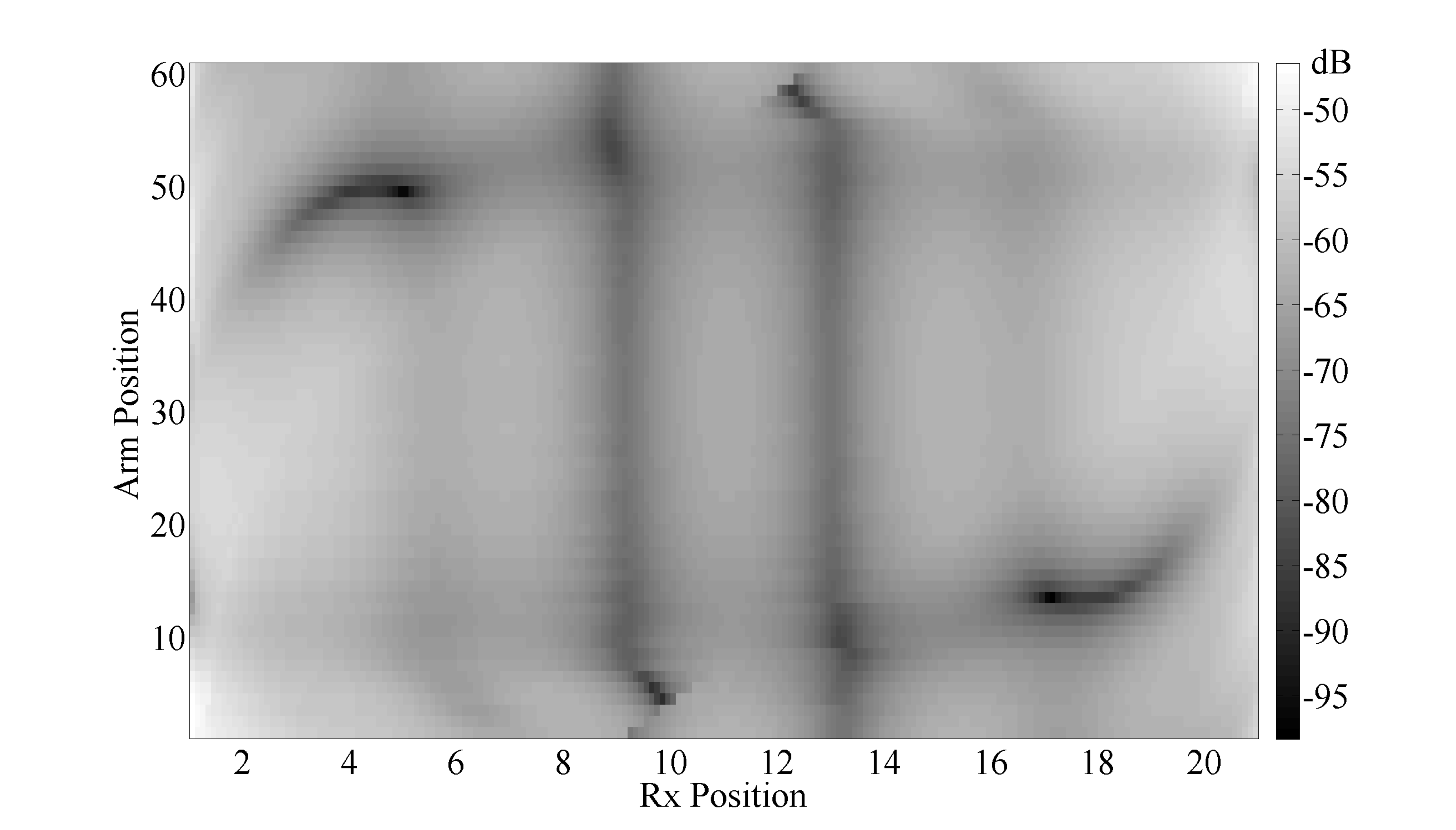}
      \caption{Variation of $S_{21}$ at different Rx positions for different arm positions}%
     \label{s21_var}
     \vspace{-8pt}
   \end{figure}

Let us consider another example of the usage of the model at $2.45$ GHz for a case when $a = 140$ mm, $b = 93.5$ mm and $r = 40$ mm. The transmitter is fixed at the central back position and $-7.3$ dBi is used as the antenna gain. The x-coordinate of the center of the left arm is kept fixed at $-(a+2r)$ and at $(a+2r)$ for the right arm. The y-coordinate of the left arm is moved from $-1.5b$ to $1.5b$ and $1.5b$ to $-1.5b$ for the right arm, at an interval of $0.05b$, simultaneously. The position where y-coordinate of the left arm is $-1.5b$ and that of the right arm is $1.5b$, is called arm position 1 and so on. Fig.~\ref{s21_var}, shows the variation of $S_{21}$ at the different receiver positions for different arm positions calculated using~(\ref{rx_pwr_elliptical_reflection}). The worst case link loss for this case is about $98$ dB whereas it is $81$ dB without considering the reflections from the arms. Hence, it is important to consider the link loss with the arms for a reliable link.

\section{Conclusions}  
An analytical model for the link loss around the torso which includes the effects of the arms was presented. The model was validated through the full-wave simulations. It was shown that the reflection of the waves from the arms at some receiver positions around the torso resulted in a higher link loss than for a case without the arms. Hence, the effects of the arms have to be considered for a proper estimation of the link loss for a reliable link between the WBAN devices placed around the torso. The input parameters needed for the model are the dimensions of the human torso and the arms. These can be obtained by measurement on the user and the link loss can be estimated for different receiver positions around the torso for any position of the arms and the transmitter.   %The model is based on the attenuation of the creeping wave on the elliptical approximation of the human's torso. A good agreement between the analytical model and the full-wave simulation is obtained. The model could be used to calculate the link loss when the devices in the body area network are located around the torso at the same level and on the opposite side. The main advantage of the model is in its simple handling for estimating the link loss in very short time when compared with the time and the memory consuming computer simulations.   

%%%%%%%%%%%%%%%%%%%%%%%%%%%%%%
  %  \addtolength{\textheight}{ -5.5cm}   % This command serves to balance the column lengths
                                  % on the last page of the document manually. It shortens
                                  % the textheight of the last page by a suitable amount.
                                  % This command does not take effect until the next page
                                  % so it should come on the page before the last. Make
                                  % sure that you do not shorten the textheight too much.   
%%%%%%%%%%%%%%%%%%%%%%%%%%%%%%%%%%%%%%%%%%%%%%%%%%%%%%%%%%%%%%%%%%%%%%%%%%%%%%%%  

%%%%%%%%%%%%%%%%%%%%%%%%%%%%%%%%%%%%%%%%%%%%%%%%%%%%%%%%%%%%%%%%%%%%%%%%%%%%%%%%

\end{document}